\documentclass[twocolumn,showpacs,floatfix,prb]{revtex4}
\usepackage[pdftex]{graphicx}
\usepackage{amssymb}
\usepackage{bm}
\usepackage{color}
\usepackage{graphicx}
\usepackage{dcolumn}
\usepackage{amssymb,amsmath,latexsym,amsbsy}
\usepackage[all]{xy}
\usepackage[dvips]{epsfig}

\newcommand{\be}{\begin{equation}}
\newcommand{\ee}{\end{equation}}
\newcommand{\bea}{\begin{eqnarray}}
\newcommand{\eea}{\end{eqnarray}}

\begin{document}


\title{Visibility recovery by strong interaction in an electronic Mach-Zehnder interferometer}


\author{Soo-Yong Lee$^1$} \email[]{phylove@postech.ac.kr}
\author{Hyun-Woo Lee$^1$}
\author{H.-S. Sim$^2$} \email[]{hssim@kaist.ac.kr}

\affiliation{$^1$Department of Physics, Pohang University of Science and Technology, Pohang, Kyungbuk 790-784, Korea}
\affiliation{$^2$Department of Physics, Korea Advanced Institute of Science and Technology, Daejeon 305-701, Korea}

\date{\today}

\begin{abstract}
We study the evolution of a single-electron packet of Lorentzian shape along an edge of the integer quantum Hall regime or in a Mach-Zehnder interferometer, considering a capacitive Coulomb interaction and using a bosonization approach. When the packet propagates along a chiral quantum Hall edge, we find that its electron density profile becomes more distorted from Lorentzian due to the generation of electron-hole excitations, as the interaction strength increases yet stays in a weak interaction regime. However, as the interaction strength becomes larger and enters a strong interaction regime, the distortion becomes weaker and eventually the Lorentzian packet shape is recovered.
The recovery of the packet shape leads to an interesting feature of the interference visibility of the symmetric Mach-Zehnder interferometer whose two arms have the same interaction strength. As the interaction strength increases, the visibility decreases from the maximum value in the weak interaction regime, and then increases to the maximum value in the strong interaction regime.
We argue that this counterintuitive result also occurs under other types of interactions.
\end{abstract}



\maketitle

\section{Introduction}


The effects of electron-electron interactions on electron interference have been recently investigated in a systematic way in experiments, by using the electronic Mach-Zehnder interferometer~\cite{Ji03Nature} realized by one-dimensional chiral edge states in the quantum Hall regime. The interactions can cause dephasing, because electrons sense  "which-path" information of other electrons through the interactions. Experiments on the interferometer have revealed nontrivial interaction-induced dephasing effects such as the so-called lobe structure~\cite{Litvin07PRB,Neder06PRL,Roulleau07PRB,Roulleau08PRL,Bieri09PRB} of the interference visibility under nonequilibrium. Different aspects of the dephasing effects have been theoretically studied in various ways of a bosonization approach,~\cite{Chalker07PRB,Neder08PRL,Levkivskyi08PRB,Levkivskyi09PRL,Gutman10PRB,Kovrizhin09PRB,Kovrizhin10PRB,Schneider11PRB} a shot-noise argument,~\cite{Levkivskyi09PRL,Youn08PRL} an inter-edge interaction model ,~\cite{Levkivskyi08PRB,Levkivskyi09PRL} and an exactly solvable model.~\cite{Kovrizhin09PRB,Kovrizhin10PRB}

Whereas most previous studies dealt with the dephasing effects in the case that electrons are continuously injected, by dc bias voltage, into the Mach-Zehnder interferometer, here we examine a simpler problem where a single isolated electron wave packet is injected to the interferometer. This situation may allow to directly investigate the dephasing of a single electron due to its interaction with the underlying Fermi sea.~\cite{Degiovanni09PRB,Lebedev11PRL} This situation can be experimentally realized by combining the interferometer with a single electron source~\cite{Keeling06PRL,Feve07Science,Kataoka11PRL} where an electron is pumped by a time dependent potential. 



In this work, we study the interaction-induced dephasing of a single electron packet moving along a chiral quantum Hall edge or through a Mach-Zehnder interferometer at filling factor $\nu = 1$. We consider a packet of Lorentzian shape and a capacitive Coulomb interaction of charging energy type. We treat the interaction, by using a bosonization method~\cite{Delft98AP} and the exactly solvable model of Kovrizhin and Chalker~\cite{Kovrizhin09PRB,Kovrizhin10PRB} that allows us to study the interferometer with the beam splitters of arbitrary transmission probability (see quantum point contacts, QPCs, in Fig.~\ref{fig1}).
When the packet propagates along the chiral edge, we find that its electron density profile becomes more distorted from Lorentzian due to the generation of electron-hole excitations, as the interaction strength increases yet stays in a weak interaction regime. However, as the strength becomes larger and enters a strong interaction regime, the distortion becomes weaker and eventually the packet shape becomes Lorentzian.
The recovery of the packet shape leads to an interesting feature of the interference visibility of the symmetric Mach-Zehnder interferometer whose two arms have the same interaction strength. As the interaction strength increases, the visibility decreases from the maximum value in the weak interaction regime, and then increases to the maximum value in the strong interaction regime. This behavior of the revival of coherence is an example~\cite{Kim09PRL} counterintuitive to the common expectation that stronger interactions may cause more dephasing.
We argue that this behavior is not specific to the capacitive interaction but can also appear under other type of interactions.


This paper is organized as follows. In Sec.~\ref{sec:setup}, we introduce the setup and the bosonization technique. In Sec.~\ref{timeevolutionofthebosonicfield}, we provide the analytical expression of the time evolution of the electron phase operator. In Sec.~\ref{Lorentzian single}, we address the dynamics of a Lorentzian packet along a quantum Hall edge. In Sec.~\ref{Oscillating current density}, we investigate the dephasing in the interferometer. In Sec.~\ref{visibility recovery}, we argue that our finding can appear in a wide class of interaction models.

\section{setup and bosonization}
\label{sec:setup}

\begin{figure}[tb]
\includegraphics[width=0.48\textwidth]{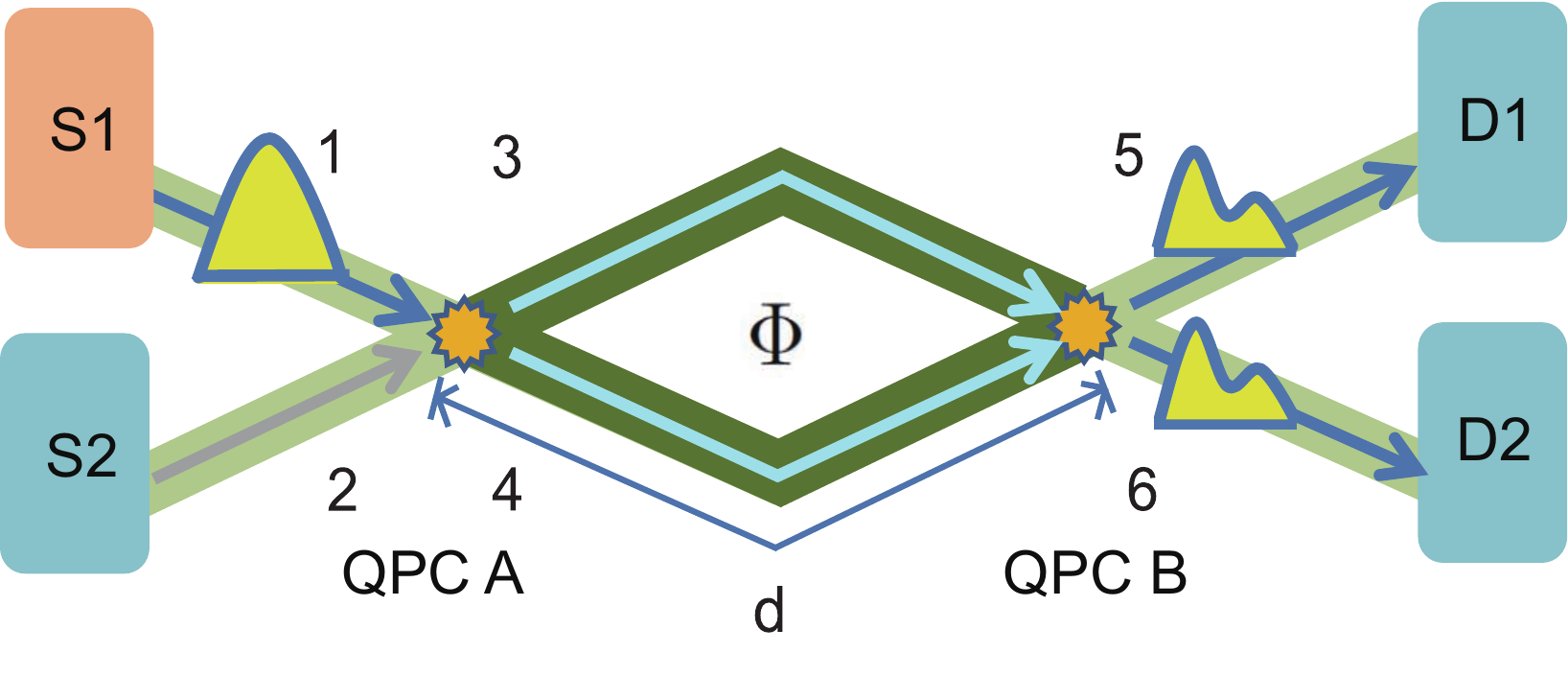}
\caption{(Color online) (a) Schematic view of an electronic Mach-Zehnder interferometer. A single electron packet is injected from Source 1 (region 1), splits at the first quantum point contact (QPC A), passes through the two arms (regions 3 and 4), and then is collected in Drain 1 (region 5) after passing the second quantum point contact (QPC B). The two arms are symmetric, having the same length $d$ and the same interaction strength, and enclose magnetic flux $\Phi$.
}
\label{fig1}
\end{figure}

The interferometer setup~\cite{Ji03Nature} is shown in Fig.~\ref{fig1}.
It consists of two sources (regions 1 and 2), two chiral interferometer arms (regions 3 and 4), and two drains (regions 5 and 6). Each arm is realized by a chiral quantum Hall edge channel at filling factor $\nu=1$, and the beam splitters by quantum point contacts (QPCs A and B). We focus on the symmetric interferometer whose two arms have the same length $d$ and the same interaction strength.

The electron field operator at coordinate $x$ in each region $i$ ($=1,2,3,4,5,6$) is denoted by $\psi_i(x)$.
For computational simplicity, we consider the situation that the total length $L$ of the system is finite but much longer than $d$, and assign coordinate as $x \in (-L/2, -d/2)$ in regions 1 and 2, $x \in (-d/2, d/2)$ in 3 and 4, and $x \in (d/2, L/2)$ in 5 and 6. The QPCs A and B are located at $x = -d/2$ and $d/2$, respectively. The scattering of the electron field operators  occurs at the QPCs as
\begin{eqnarray}
\left(\begin{array}{c} \hat{\psi}_3(x) \\ \hat{\psi}_4(x) \end{array} \right)_{x=-\frac{d}{2}+0} & = & {{\cal S}_\textrm{A}} \left( \begin{array}{c} \hat{\psi}_1(x) \\ \hat{\psi}_2(x)  \end{array} \right)_{x=-\frac{d}{2}-0}, \nonumber \\
\left( \begin{array}{c} \hat{\psi}_5(x) \\ \hat{\psi}_6(x) \end{array} \right)_{x=\frac{d}{2}+0} & = & {{\cal S}_\textrm{B}}{{\cal S}_{\Phi}}
\left( \begin{array}{c} \hat{\psi}_3(x) \\ \hat{\psi}_4(x)  \end{array} \right)_{x=\frac{d}{2}-0}, \nonumber
\end{eqnarray}
where ${{\cal S}_j} = \left (\begin{array}{cc}{r_j}&{it_j}\\{it_j}&{r_j}\end{array} \right)$ is the scattering matrix at QPC $j$ ($= \textrm{A}, \textrm{B}$), ${\cal S}_j {\cal S}_j^\dagger =  {\cal S}_j^\dagger {\cal S}_j  = I$,  and $r_j$ ($t_j$) is the reflection (transmission) coefficient of QPC $j$; we choose, for simplicity, $t_j$ and $r_j$ as real. The effect of the magnetic flux $\Phi$ ($=\Phi_{3}-\Phi_{4})$ enclosed by the two arms (regions 3 and 4) is described by ${{\cal S}_\Phi} = \left (\begin{array}{cc}{e^{i \Phi_{3}}}&{0}\\{0}&{e^{i \Phi_{4}}}\end{array} \right)$.

Below, we describe the bosonization approach for the low energy regime of the system. The total Hamiltonian is decomposed into the kinetic and interaction parts, ${\hat{H}_{\mathrm{tot}}}={\hat{H}_{\mathrm{kin}}}+{\hat{H}_{\mathrm{int}}}$. The kinetic part has the linear form of
$(\hbar v_F / i) \sum_{j =
1}^{6} \int dx : \hat{\psi}^{\dagger}_{j}(x) \partial_x \hat{\psi}_{j}(x):$,
\be\label{kinetic_Hamiltonian}{\hat{H}_\mathrm{kin}}= v_F \sum_{l = u,d} \int^{L/2}_{-L/2} dx : \hat{\psi}^{\dagger}_{l}(x)
\frac{\hbar}{i} \partial_x \hat{\psi}_{l}(x): \ee
where $v_F$ is the Fermi velocity, $: \cdots :$ stands for the normal ordering, and we introduced
operators $\hat{\psi}_u$ and $\hat{\psi}_d$,
\begin{equation} \left( \begin{array}{c}
\hat{\psi}_u(x) \\ \hat{\psi}_d(x) \end{array} \right) = \left\{
\begin{array}{ll} \left( \begin{array}{c} \hat{\psi}_3(x) \\
\hat{\psi}_4(x) \end{array} \right) & \textrm{for $-\frac{d}{2}< x
< \frac{d}{2}$} \\  {{\cal S}_a}\left( \begin{array}{c}
\hat{\psi}_1(x) \\ \hat{\psi}_2(x) \end{array} \right) &
\textrm{for $-\frac{L}{2} < x < -\frac{d}{2}$} \\{{\cal
S}_{\Phi}}^{\dagger}  {{\cal
S}_b}^{\dagger}\left( \begin{array}{c} \hat{\psi}_5(x) \\
\hat{\psi}_6(x) \end{array} \right)  & \textrm{for $\frac{d}{2} <
x < \frac{L}{2}$} \end{array} \nonumber \right. .\end{equation}
$\hat{\psi}_u$ and $\hat{\psi}_d$ are defined over the entire range of $-L/2 < x <L/2$, and continuous at $x=\pm d/2$. They capture the effects of the QPC's. On the other hand, the electron-electron interactions in the two interferometer arms are described, as in previous studies ~\cite{Sukhorukov07PRL,Neder08PRL,Youn08PRL,Schneider11PRB,Kovrizhin09PRB,Kovrizhin10PRB}, by a capacitive interaction of the charging energy type,
\be
\label{interaction_Hamiltonian} {\hat{H}_{\mathrm{int}}}=
\frac{1}{2} \frac{g v_F \hbar}{d} \sum_{l = u,d}
\int^{d/2}_{-d/2} dx dx' :\hat{\rho}_{l}(x)
\hat{\rho}_{l}(x'): \ee where $g$ is the
dimensionless interaction strength and
$\hat{\rho}_l(x)=:\hat{\psi}_l^{\dagger}(x)\hat{\psi}_l(x):$ is
the electron density operator in channel $l$.
We ignore the interactions in the sources and drains (regions 1, 2, 5, 6).

From the form of $\hat{H}_\textrm{tot}$, one notices that $\hat{\psi}_u$ and $\hat{\psi}_d$ are completely decoupled from each other, acting as the ``eigenchannels'' of $\hat{H}_{\mathrm{tot}}$. This simplifies the analysis of $\hat{H}_{\mathrm{tot}}$.
We impose the periodic boundary conditions
$\hat{\psi}_{l=u,d} (-L/2)=\hat{\psi}_l (L/2)$ onto each channel, and define the electron annihilation
operators $\hat{c}_{l,k}$ and the electron density operators  $\hat{\rho}_l(k)$ of channel $l$ in the momentum space by
$\hat{\psi}_l(x)=\frac{1}{\sqrt{L}} \sum_k e^{ikx} \hat{c}_{l,k}$ and $\hat{\rho}_l(x)=\frac{1}{L} \sum_k e^{ikx}
\hat{\rho}_l(k)$, where $k = 2\pi n /L$ and $n\in\mathbb{Z}$. $\hat{\rho}_l$ satisfies the commutation rules \cite{Delft98AP} of $[\rho_l
(q), \rho_{l'} (-q')]=\frac{qL}{2 \pi} \delta_{l,l'} \delta_{q,q'}$ and $[\rho_l (x), \rho_{l'}
(x')]=\frac{i}{2 \pi} \partial_{x'}{\delta(x-x')} \delta_{l,l'}$.
It is decomposed \cite{Delft98AP} into $\hat{\rho}_l(x)=\frac{1}{2 \pi} \partial_x
\hat{\phi}_l(x)+\frac{\hat{N}_l}{L}$. $\hat{N}_l \equiv
\int^{L/2}_{-L/2} dx \hat{\rho}_l(x)$ is the zero-mode operator counting the number of electrons in channel $l$ and $\hat{\phi}_l(x)$ is the bosonic operator describing the plasmon excitations of channel $l$,
\be\hat{\phi}_l(x)=\frac{2 \pi}{L} \sum_{q \neq 0} \frac{1}{iq} e^{iqx} e^{-|q|a/2} \hat{\rho}_l(q),\ee  where $a$ is an infinitesimal positive real constant introduced to regularize divergent sums.
The bosonic operator $\hat{\phi}_l(x)$ is related to the electronic field $\hat{\psi}_l$,
\be
\label{fermion-boson-relation}\hat{\psi}_l(x)=\frac{1}{\sqrt{2 \pi
a}}\hat{F}_l e^{i \frac{1}{L} \hat {N}_l x} e^{i
\hat{\phi}_l(x)}, \ee where $\hat{F}_l$ is the Klein operator that
reduces the eigenvalue of $\hat{N}_l$ by $1$. From Eq.~\eqref{fermion-boson-relation}, one can interpret $\hat{\phi}_l(x)$ as the electron phase operator. Then $\hat{H}_{\mathrm{kin}}$ is
bosonized~\cite{Delft98AP}
\be
\label{density_kinetic_Hamiltonian}{\hat{H}_{\mathrm{kin}}}=
\frac{v_F h}{2} \sum_{l =u,d}  [\int^{L/2}_{-L/2} dx :
\hat{\rho}_{l}(x)^2:+\frac{\hat{N}_l}{L}],\ee
therefore, the total Hamiltonian $\hat{H}_{\mathrm{tot}}$
is expressed in terms of the bosonic operators $\hat{\rho}_l(x)$
and $\hat{N}_l$.

\section{Time evolution of the phase operator}
\label{timeevolutionofthebosonicfield}

In this section, we analytically study the time evolution of the bosonic phase operator $\hat{\phi}_l(x)$. We note that
the introduction of the ``eigenchannels'' $\hat{\psi}_u$ and $\hat{\psi}_d$ in Eq.~\eqref{kinetic_Hamiltonian} allows the analytic study; a similar problem has been studied by Kovrizhin and Chalker.~\cite{Kovrizhin09PRB,Kovrizhin10PRB}

The time dependence of $\hat{\phi}_l(x)$ is written as
\be\label{bosonicfields}
\hat{\phi}_{l}(x,t)=\frac{2 \pi}{L}\sum_{q \neq 0 } \frac{1}{i q}
e^{iqx}e^{-|q|a/2}\hat{\rho}_{l}(q,t).\ee
{Here $x=0$ denotes the center of arm and $t=0$ stands for an initial time.} After some algebra, one finds that
$\hat{\phi}_l(x,t)$ satisfies the equation of
motion,
\be \label{equationofmotion} [\partial_t+v_F
\partial_x] \hat{\phi}_l(x,t)=-\frac{g v_F}{2 \pi d}
[\hat{\phi}_l(\frac{d}{2},t)-\hat{\phi}_l(-\frac{d}{2},t)+\frac{d}{L}\hat{N}_l]
\ee
for $-d/2\leq x \leq d/2$, and $[\partial_t+v_F \partial_x]
\hat{\phi}_l(x,t)=0$ otherwise.
In the non-interacting case $g=0$, $\hat{\phi}_{l}(x,t)$
satisfies the zeroth-order solution of
$\hat{\phi}^{(0)}_l(x,t)=\hat{\phi}_{l}(x-v_F t,0)$. We note that the time dependence of the zero-mode $\hat{N}_l$ is neglected because of $L \rightarrow \infty$.

In the presence of the interaction with nonzero $g$, $\hat{\phi}_{l}(x,t)$ can be
expanded as
$\hat{\phi}_l(x,t)=\hat{\phi}^{(0)}_l(x,t)+\hat{\phi}^{(1)}_l(x,t)+\hat{\phi}^{(2)}_l(x,t)
\cdots$ with respect to the order of $g$.
When $x \in (-L/2, - d/2)$ or $x \in (d/2, L/2)$, $\hat{\phi}_l^{(n)}
(x,t)$ satisfies $(\partial_t+v_F \partial_x)\hat{\phi}_l^{(n)}
(x,t)=0$ for all $n$. In the case of $x \in (-d/2, d/2)$, we derive, from Eq.~\eqref{equationofmotion}, the recurrence relation between
$\hat{\phi}^{(n+1)}(x,t)$ and $\hat{\phi}^{(n)}(x,t)$ for $n \geq 1$,
\be \label{recursiveeom}
[\partial_t+v_F \partial_x] \hat{\phi}^{(n+1)}_l(x,t)=-\frac{v_F}{d}
\frac{g}{2 \pi}
[\hat{\phi}^{(n)}_l(\frac{d}{2},t)-\hat{\phi}^{(n)}_l(-\frac{d}{2},t)].\ee
Once $\hat{\phi}_l^{(1)}(x,t)$ is obtained, all
$\hat{\phi}_l^{(n \geq 2)}$'s can be recursively obtained from Eq.
(\ref{recursiveeom}).

To obtain the first-order solution of $\hat{\phi}_l^{(1)}(x,t)$, we first evaluate
$\hat{\rho}_l(q,t)=e^{i\hat{H}_{\mathrm{tot}}t/\hbar}
\hat{\rho}_l(q)
e^{-i\hat{H}_{\mathrm{tot}}t/\hbar}=\hat{\rho}^{(0)}_l(q,t)+\hat{\rho}^{(1)}_l(q,t)+\hat{\rho}^{(2)}_l(q,t)+\cdots
$ by using the Baker-Haussdorff lemma of $\hat{\rho}_l(q,t)=\hat{\rho}_l(q)+\frac{it}{\hbar}[\hat{H}_{\mathrm{tot}},\hat{\rho}_l(q)]+\frac{1}{2!} (\frac{it}{\hbar})^2[\hat{H}_{\mathrm{tot}},[\hat{H}_{\mathrm{tot}},\hat{\rho}_l(q)]]+\cdots$. One can easily verify that the zeroth-order contribution of
$\hat{\rho}^{(0)}_l(q,t)$ is given by $e^{-iqv_F t}
\hat{\rho}_l(q)$. The evaluation of the first-order contribution of
$\hat{\rho}^{(1)}_l(q,t)$ is rather tedious and given in Appendix
\ref{Additional_density_and_field_operator}. One obtains $\hat{\phi}^{(1)}_l(q,t)$ by inserting
$\hat{\rho}^{(1)}_l(q,t)$ into Eq. (\ref{bosonicfields}), and
$\hat{\phi}^{(n\geq 2)}_l(q,t)$ by using Eq. (\ref{recursiveeom}).

Then, we obtain
$\delta \hat{\phi}_l(x,t) \equiv \hat{\phi}_l(x,t)-\hat{\phi}^{(0)}_l(x,t)$ as
 \begin{equation} \label{delta_phiqt}\delta \hat{\phi}_l(x,t) = -\sum_{q} K(q;x,t) \hat{\rho}_l(q) e^{iq(x-v_F t)}. \end{equation}
For the case of $x>d/2$ (regions 5 and 6) and $t>
\frac{x+d/2}{v_F}$ (propagation time from the left end of the arms at $-d/2$ to $x$), we find that the kernel $K(q;x,t)$ reduces to the form of $K(q)$, which is independent of $x$ and $t$,
\be\label{chargingkernel}K(q)=\frac{d}{L}\frac{g
(\frac{\sin{qd/2}}{qd/2})^2}{1+\frac{g}{2 \pi}
\frac{\sin{qd/2}}{qd/2} e^{iqd/2}},\ee
and that $\delta \hat{\phi}_l(x,t)$ reduces to $\delta \hat{\phi}_l(x-v_F t,0)$. $K(q)$ shows the transition amplitude of an electron with momentum difference $q$ by e-e interaction. Thus, this analytic expression is very useful for understanding single electron dynamics even in a strong interaction regime as below. We note that analytic $K(q)$ agrees with the kernel obtained in Refs. \onlinecite{Kovrizhin09PRB} and \onlinecite{Kovrizhin10PRB}.



\begin{figure}[bt]
\includegraphics[width=0.47\textwidth]{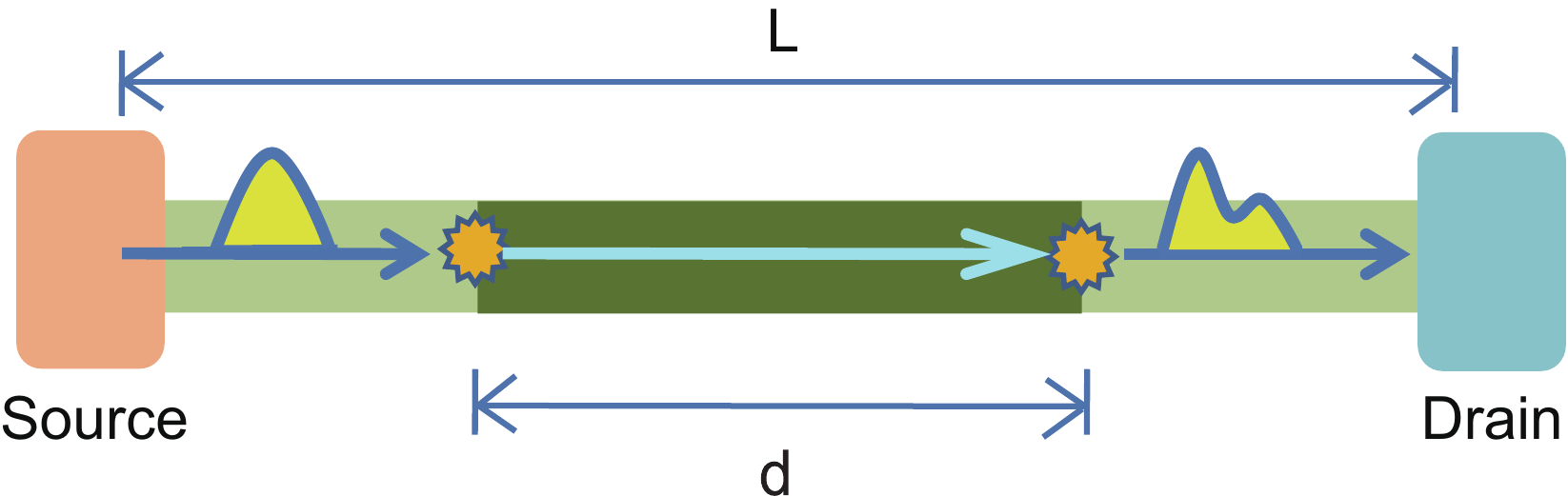}
\caption{(Color online) Schematic view of the single chiral channel where the electron-electron interaction is present only within the dark region of length $d$.
}
\label{fig2}
\end{figure}

\section{Propagation of a Lorentzian packet along a chiral channel}
\label{Lorentzian single}

We first investigate the propagation of a single
electron wave packet along a chiral channel at $\nu = 1$. We consider the situation in Fig.
\ref{fig2} that the capacitive Coulomb interaction is present only within a region of length $d$. The Hamiltonian of the channel is
\begin{eqnarray}
\hat{H}_{\mathrm{ch}} & = & v_F
\int^{\frac{L}{2}}_{-\frac{L}{2}} dx : \hat{\psi}^{\dagger}(x)
\frac{\hbar}{i} \partial_x \hat{\psi}(x): \nonumber \\
& + &
 \frac{g v_F
\hbar}{2d} \int^{\frac{d}{2}}_{-\frac{d}{2}} dx dx' : \hat{\rho}(x)
\hat{\rho}(x'):,  \nonumber \end{eqnarray}
where $\hat{\rho}(x)=:
\hat{\psi}^{\dagger}(x) \hat{\psi}(x):$.
This describes the Mach-Zehnder interferometer with $r_\textrm{A} = r_\textrm{B} = 0$. We will examine in this section how the wave packet is distorted as it passes through the interaction
region. The insights obtained in this section will be useful for
understanding the the interaction effect on the interference visibility
of the single wave packet in the Mach-Zehnder interferometer,
which is the subject of the next section.

We confine ourselves to one particular type of a wave packet, a
Lorentzian packet. Its shape facilitates the
analytic calculation considerably. The Lorentzian packet also has
practical merits as it can be created by a Lorentzian voltage
pulse with {\it{minimal}} noise. \cite{Keeling06PRL} A Lorentzian packet created on top of the filled Fermi sea $|F\rangle$ is expressed as $ | {\Psi}_{\xi}(X)
\rangle = \int dx f_{\xi}(x;X) \hat{\psi}^{\dagger}(x) |F \rangle$
where $ f_{\xi}(x;X)=\sqrt{\frac{\xi}{\pi}}\frac{i}{x-X+i\xi}$.
The electron density $\langle {\Psi}_{\xi}(X)| \hat{\rho}(x) |
{\Psi}_{\xi}(X) \rangle$ generated by the packet has the
Lorentzian profile of $ \frac{\xi}{\pi} \frac{1}{(x-X)^2+\xi^2}$ with packet center
$x=X$ and width $\xi$. In the momentum space, it is written as
\begin{eqnarray} \label{noneqstate} |\Psi_{\xi}(X)\rangle= {\sqrt\frac{4 \pi \xi }{L}}\sum_{k>0} \hat{c}^{\dagger}_{k} e^{i k (-X+i\xi)}
|F\rangle.\end{eqnarray}Note that the summation over $k$ runs only
over positive $k$ (above the Fermi sea).

We study the time evolution of the packet whose center is initially located at $X \ll -d / 2 - \xi$ in the left side of the interaction region. As time goes on, it moves to the right. We calculate the expectation value of the density operator at position $Y \gg d/2$ in the right side of the interaction region, $\rho_\textrm{ch}(Y,t)=\langle \Psi_{\xi}(X)  |
\hat{\rho}(Y,t) |\Psi_{\xi}(X)\rangle$.
The time dependence of the density operator is decomposed as $\hat{\rho}_\textrm{ch}(x,t) =  \hat{\rho} (x-v_F t,0)+\delta\hat{\rho} (x,t)$. The first term of $\hat{\rho} (x-v_F t,0)$
is the trivial density of the non-interacting case that preserves the original Lorentzian shape, while the second term of $\delta\hat{\rho} (x,t)=
\frac{1}{2 \pi}\partial_x \delta\hat{\phi} (x,t)$ describes the distortion due to the interaction.
From Eq. (\ref{delta_phiqt}), we obtain the distortion part $\delta
{\rho}_\textrm{ch} (Y,t)=\langle \Psi_{\xi}(X)  | \delta\hat{\rho}(Y,t)
|\Psi_{\xi}(X)\rangle$ (See Appendix \ref{Additional
density operator}) \bea \label{Additional density operator eq}
\delta \rho(Y,t)=\frac{1}{2 \pi}\sum_{q} -iq K(q) e^{-|q|\xi}
e^{iq(Y-X-v_Ft)}.\eea  By using the
analytic expression of $K(q)$ in Eq.
(\ref{chargingkernel}), the electron density profile $\rho_\textrm{ch}(Y,t)$
is easily evaluated. Note that $\int dY
\rho_\textrm{ch} (Y,t) =1$ because of charge conservation.


\begin{figure}[tb]
\includegraphics[width=0.47\textwidth]{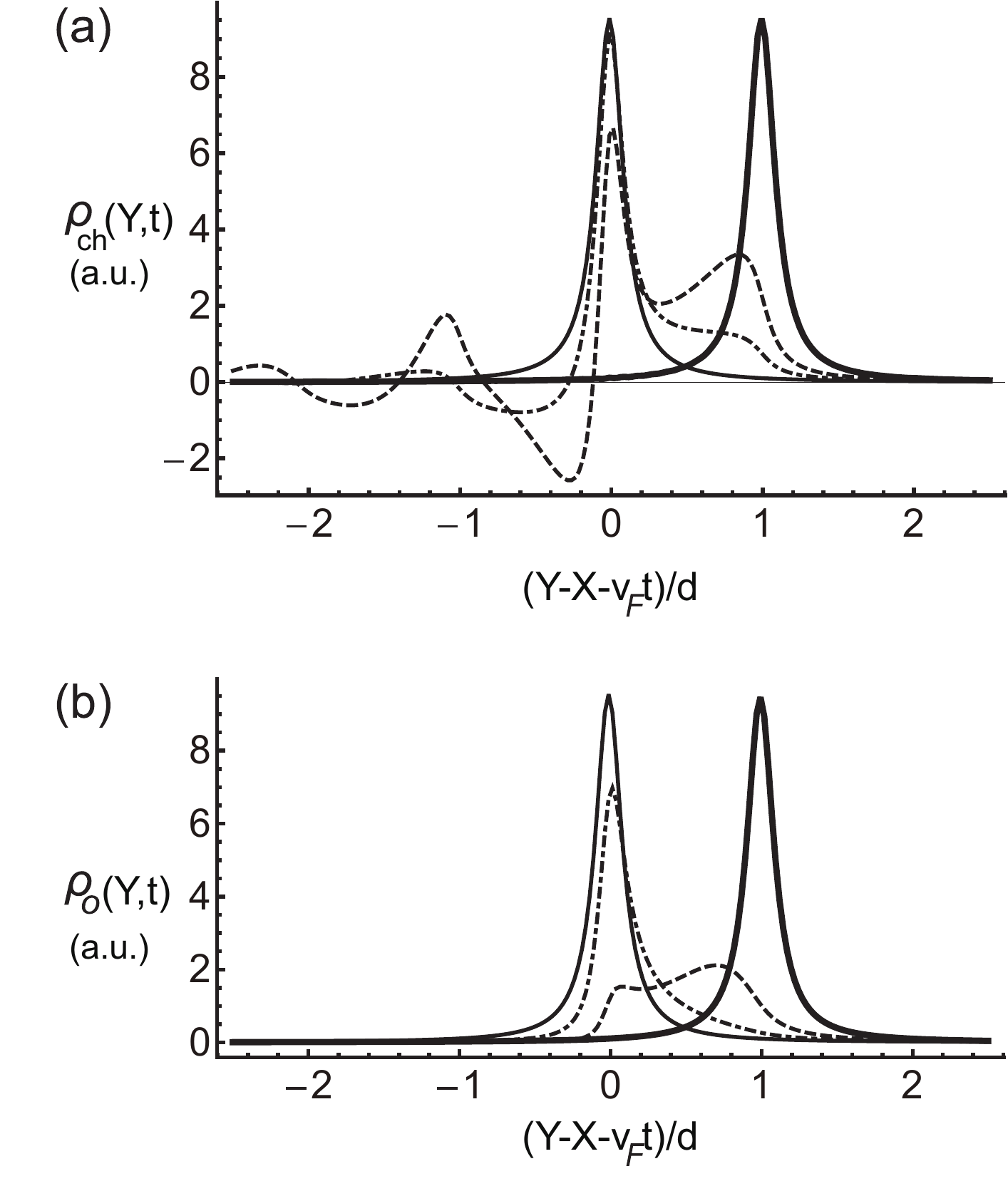}
\caption{(a) Electron density profile in (a) a chiral channel and in (b) a Mach-Zehnder interferometer, after a Lorentzian packet passes through the interaction region. In (b), only the magnetic-flux-dependent part $\rho_{o}$ of the density profile is depicted with $t_A^2=t_B^2=0.5$. The profile is shown for different values of the interaction strength, $g/(2\pi)=0$ (solid line), $0.5$ (dot-dashed), $2$ (dashed), and $1000$ (thick solid). We choose $d/L=0.05$ and $2 \pi \xi=2d/3$.}\label{fig3}\end{figure}

The result is shown in Fig. \ref{fig3}(a) as
a function of $g$. As $g$ grows, 
the electron density profile more deviates from the Lorentzian profile, because of the creation of particle-hole pair excitations due to the interaction.
Interestingly, in the strong interaction limit of $g \to \infty$, the packet recovers its original Lorentzian
profile but with the center shifted by the extra distance of $d$. Mathematically, this feature arises since $K(q) \to \frac{2 \pi i}{L} (e^{-iqd}-1)/q$ as $g \rightarrow \infty$ [see Eq. (\ref{chargingkernel})], which yields
$\rho_\textrm{ch} (Y,t)=\frac{1}{L}\sum_{q} e^{-|q|\xi} e^{i q (
Y-X-v_F t-d)} = \rho_\textrm{ch} (Y-d-v_F t,0)$.
Physically, this feature may be understood as follows. In the $g \rightarrow \infty$ limit, the strong interaction suppresses the charge fluctuations in the region of $- d/2 <x< d/2$. Then as soon as charges are injected to the interaction region from the left at $x=-d/2$, the exactly same amount of charges are ejected from the interaction region to the right at $x=d/2$, because of the chiral property. Otherwise the total charge in the interaction region should be modified, which is energetically very costly. The balance between the injected charge amount and the ejected charge amount should be maintained at each time instance. This explains the shift of the Lorentzian packet by distance $d$ without distortion. We expect that this feature will also occur under other types of electron interactions in the strong interaction limit where charge fluctuations are suppressed in the interaction region.

\section{Lorentzian packet in a Mach-Zehnder interferometer}
\label{Oscillating current density}

In this section, we study the interference of a single Lorentzian packet in a Mach-Zehnder interferometer at $\nu = 1$. The Lorentzian packet $| \Psi_{1,\xi} (X) \rangle  = \int dx f_{\xi}(x;X) \hat{\psi}^{\dagger}_1(x) |F \rangle$ is incoming from region $1$, and detected in region 5; see Fig. \ref{fig1}(a).

The current density operator $\hat{I}_5 =ev_F \hat{\rho}_5$ of region 5 ($x>d/2$) satisfies $\partial_x\hat{I}_5(x,t)=-[e\hat{\rho}_5(x,t),\hat{H}_{\mathrm{tot}}]/(i \hbar)$, where $\hat{\rho}_5 \equiv :\hat{\psi}_5^{\dagger}(x,t)\hat{\psi}_5(x,t):$. $\hat{I}_5$ is expressed in terms of $\hat{\psi}_u$ and $\hat{\psi}_d$ using Eq.~\eqref{kinetic_Hamiltonian}, and decomposed into $\hat{I}_5(x,t)=\hat{I}_{5,n}(x,t)+\hat{I}_{5,o}(x,t)$. Here $\hat{I}_{5,n}(x,t)=ev_F :r^2_B \hat{\psi}^{\dagger}_u (x,t) \hat{\psi}_u (x,t)+t^2_B \hat{\psi}^{\dagger}_d (x,t) \hat{\psi}_d (x,t) :$ is the direct term independent of the magnetic flux $\Phi$, while $\hat{I}_{5,o}(x,t)= e v_F \mathrm{Re} [:2 i r_B t_B \hat{\psi}^{\dagger}_u(x,t) \hat{\psi}_d(x,t) : e^{- i \Phi}]$ is the interference term. Accordingly, the current density $I_5(Y,t)= \langle \Psi_{1,\xi}(X) | \hat{I}_5 (Y,t) | \Psi_{1,\xi}(X) \rangle$ in region 5 is decomposed into $I_{5,n}(Y,t)$ and $I_{5,o}(Y,t)$.
We find the direct part of  $I_5$ as
\bea I_{5,n}(Y,t) &=& e v_F [r^2_B \langle \Psi_{1,\xi}(X)| \rho_u(Y,t) | \Psi_{1,\xi}(X) \rangle\nonumber\\ & &+ t^2_B \langle \Psi_{1,\xi}(X)| \rho_d(Y,t) | \Psi_{1,\xi}(X) \rangle ]\nonumber\\
 &=& ev_F (r^2_A r^2_B + t^2_A t^2_B) \rho_\textrm{ch} (Y,t), \eea
where $\rho_\textrm{ch}(Y,t)$ is the electron density profile in the chiral channel discussed in Sec.~\ref{Lorentzian single}. The interference part is
\bea \label{psirelaton}I_{5,o} (Y,t) 
& =&-2 ev_F r_A t_A r_B t_B \rho_o (Y,t) \cos \Phi ,\eea
where $ \rho_o(Y,t) \equiv \langle \Psi_{u,\xi}(X) |:\hat{\psi}^{\dagger}_u(Y,t) \hat{\psi}_d(Y,t): | \Psi_{d,\xi} (X)\rangle$ and $|\Psi_{1,\xi}(X) \rangle=r_A |\Psi_{u,\xi}(X) \rangle + it_A |\Psi_{d,\xi}(X) \rangle$ from Eq.~\eqref{kinetic_Hamiltonian}. 
By using Eq. (\ref{psirelaton}) and the fact that $\hat{\psi}_u$ and $\hat{\psi}_d$ are dynamically decoupled, one obtains {$\rho_o(Y,t)= \chi^*_u(Y,t) \chi_d(Y,t)$, where \be \label{chi}\chi_l(Y,t)= \int dx'\langle F| \hat{\psi}_l(Y,t) \hat{\psi}^{\dagger}_{l}(x',0)|F \rangle f_{\xi}(x';X)\ee for $Y\gg d/2$ (regions 5, 6) and $t>\frac{Y+d/2}{v_F}$ (propagation time from $x=-d/2$ to $Y$).} 
By using the bosonization technique, we evaluate $ \rho_o(Y,t)$ as (see Appendix \ref{bosoncorrelator})
\bea \label{densityprofileo} \rho_{o}(Y,t)&=& \frac{\xi}{\pi} \frac{e^{2 \mathrm{Im}[\sum_{q>0}  K(q) e^{-q \xi} e^{iq(Y-X-v_F t)}]}}{(Y-X-v_F t)^2+\xi^2}.\eea
Using Eq. (\ref{chargingkernel}), one computes $\rho_o(Y,t)$.

The result of $\rho_o(Y,t)$ is shown in Fig. \ref{fig3}(b) for various values of $g$. In the non-interacting case of $g=0$, $\rho_o(y,t)$ has the Lorenzian shape. As $g$ increases, $\rho_o(Y,t)$ deviates from the Lorentzian profile due to particle-hole excitations by the interaction. $\rho_o(Y,t)$ becomes to recover its original Lorentzian shape but with the center shifted by $d$, as $g$ further increases (beyond about $4 \pi$) and enters into the strong-interaction limit of $g \to \infty$. This feature has the same origin with the corresponding effect in the single chiral channel discussed in the last section. 



\begin{figure}[tb]
\includegraphics[width=0.47\textwidth]{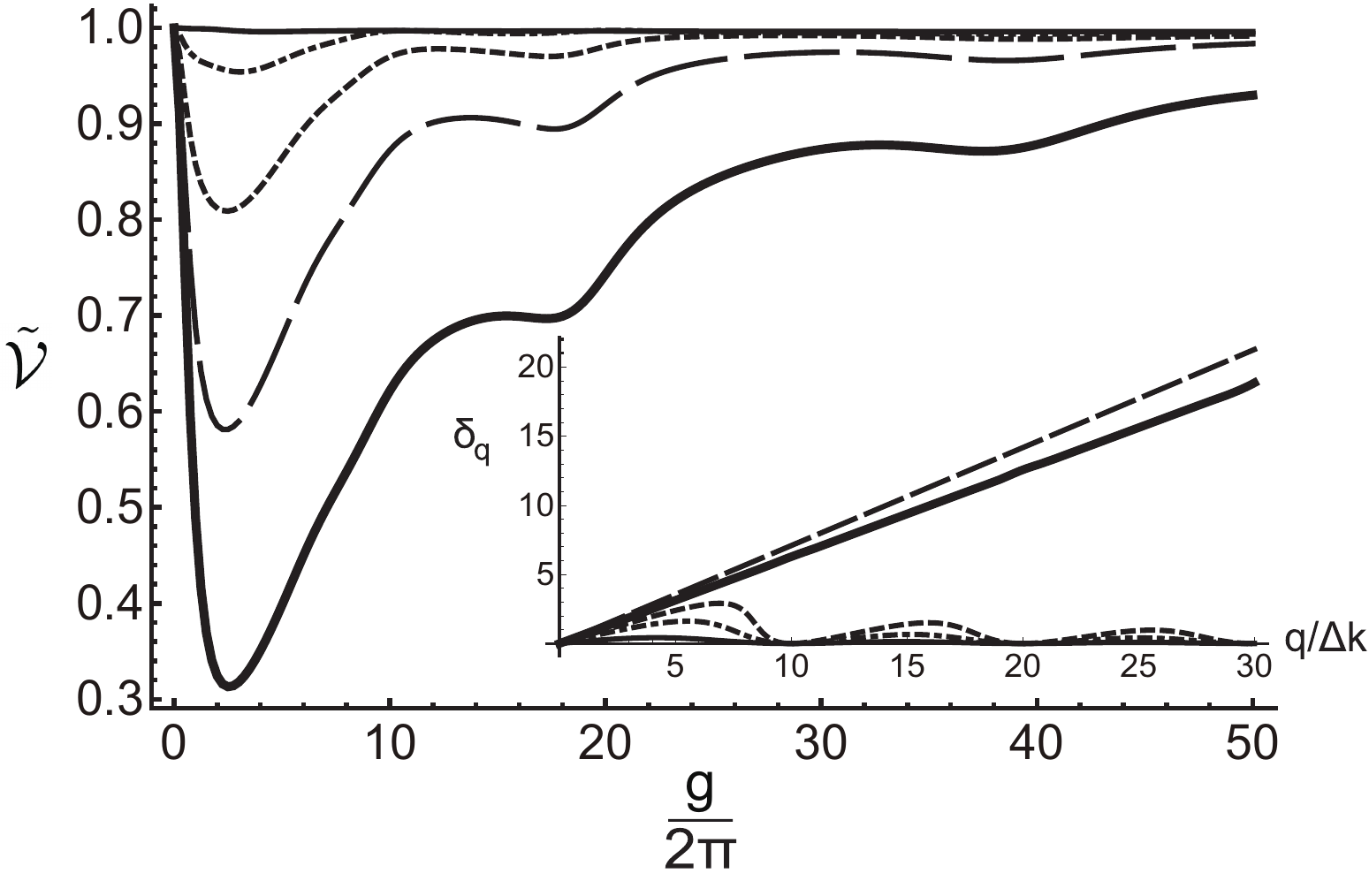}
\caption{Normalized interference visibility $\tilde{\mathcal{V}}$ as a function of $g$.
We choose the packet widths of $2 \pi\xi/d= 5$ (solid line), 2 (dot-dashed), 1 (dashed),  1/2 (long-dashed), and 1/5 (thick solid), and use $d/L=0.05$. Inset: $\delta_q$ as a function of $q$, for the capacitive interaction with $g/2\pi= 2$ (solid line), 10 (dot-dashed), 25 (dashed), 1000 (long-dashed), and for the regularized Coulomb interaction of $V_{r1}(x,x')= \frac{g}{d} b/\sqrt{(x-x')^2+b^2}$ and $V_{r2}(x,x') = \frac{g}{d} \exp{[-|x-x'|/b]}$ with $b=0.1d$, {$d/L=0.1$} and $g/2 \pi = 1000$ (two thick solid lines); the two thick solid lines for $V_{r1}$ and $V_{r2}$ almost overlap with each other and appear as a single line. This result of the linear dispersions of $\delta_q$ provides the clue that the visibility recovery also appears in the strong interaction regime of the regularized Coulomb interactions of $V_{r1}$ and $V_{r2}$.
}\label{fig4}\end{figure}

We investigate the implication of the above interesting feature on the interference visibility.
We compute the total charge transmission $Q_5(\Phi)$ to drain $1$ (region 5). $Q_5(\Phi)$ is decomposed into the flux-independent part $Q_{5,n}=\int dt I_{5,n}(Y,t)$ and the flux-dependent part $Q_{5,o}(\Phi)=\int dt I_{5,o}(Y,t;\Phi)=Q^0_{5,o} \cos \Phi$, where $Q^0_{5,o}$ is the oscillation amplitude of $Q_{5,o}(\Phi)$. Note that both $Q_{5,n}$ and $Q_{5,o}$ are independent of $Y$. The evaluation of $Q_{5,n}$ is straight forward, $Q_{5,n}=e(r^2_A r^2_B+t^2_A t^2_B)$, since $ v_F \int dt \rho_\textrm{ch}(Y,t) = 1$ due to the charge conservation. On the other hand,
$Q_{5,o}^0=2 e r_A t_A r_B t_B v_F \int dt \rho_o(Y,t)$ needs to be explicitly evaluated. In Fig. \ref{fig4}, we show the visibility $\mathcal{V} \equiv (Q_{5, \textrm{max}} -Q_{5, \textrm{min}})/(Q_{5, \textrm{max}} +Q_{5, \textrm{min}}) = Q_{5,o}^{0}/Q_{5,n}$ of the charge transmission as a function of $g/2 \pi$, where $Q_{5, \textrm{max(min)}}$ is the maximum (minimum) value of $Q_5(\Phi)$. 
$\mathcal{V}$ depends on $r_A$, $t_A$, $r_B$, $t_B$ through the combination of $\frac{2 r_A t_A r_B t_B}{r_A^2 r_B^2+ t_A^2 t_B^2}$, thus the normalized visibility $\tilde{\mathcal{V}} \equiv \mathcal{V} / [\frac{2 r_A t_A r_B t_B}{r_A^2 r_B^2+ t_A^2 t_B^2}]$ is plotted instead. In the non-interacting limit of $g \rightarrow 0$, $\tilde{\mathcal{V}}$ becomes $1$. As $g$ grows, $\tilde{\mathcal{V}}$ decreases, implying the dephasing induced by the interaction. However, as $g$ further increases beyond $\sim 4 \pi$, $\tilde{\mathcal{V}}$ becomes larger
and revives, approaching to the maximum value of $1$ in the strong interaction limit of $g \rightarrow \infty$.
Namely, the interference visibility is not a monotonically decaying function of the interaction strength, which is in contrast to the conventional expectation that stronger interactions cause more severe dephasing. The revival of the electron coherence in the strong interaction limit is the consequence of the restored density profile of $\rho_o(Y,t)$ along the chiral channel; see Fig. \ref{fig3}(b) and Sec. \ref{Lorentzian single}.

Note that Fig. \ref{fig4} also shows that $\tilde{\mathcal{V}}$ increases
as the packet width $\xi$ increases. This is natural, because larger $\xi$ implies smaller excitation energy.

\section{Discussion and conclusion}
\label{visibility recovery}


The origin of the revival of the visibility in the strong interaction regime can be understood from the suppression of particle-hole creation in the interaction regions.
This implies that the revival can also occur in the strong interaction regime of the other types of Coulomb interactions. We find numerically that the visibility revival indeed occurs in the cases of the regularized Coulomb interaction $V_{r1}(x,x')= g b / (d \sqrt{(x-x')^2+b^2})$ and the exponentially screened interaction $V_{r2}(x,x') = (g/d) \exp{[-|x-x'|/b]}$ (results not shown here). 

We discuss the revival of the visibility in another way, based on the form of the kernel $K(q)$ in { Eq. (\ref{delta_phiqt})}. For general types of electron interaction, the kernel has the form \cite{Kovrizhin09PRB,Kovrizhin10PRB}  of $-iqK(q) = 2 \pi (e^{-i \delta_q}-1)/L$. $\delta_q$ is the phase that the bosonic field $\hat{\phi}(q)$ acquires in the middle region of length $d$ and interaction strength $g$. In general, $\delta_q$ is nonlinear in $q$. In this case, there occurs dephasing, i.e., visibility reduction, because of the phase randomization in interaction-induced scattering processes between momentum states. On the other hand, there is no dephasing (i.e., no phase randomization) in the case that $\delta_q$ is linear in $q$. For example, in the case of the short-range interaction of $V(x,x') \propto \delta(x-x')$, one finds $\delta_q \propto q$ and no dephasing; in the case of the short-range interaction, the only effect of the interaction is the shift of the propagation velocity. In the cases of the capacitive interaction, the regularized Coulomb interaction $V_{r1}$, and the exponentially screened interaction $V_{r2}$, $\delta_q$ becomes proportional to $q$ in the strong interaction limit (see the inset of Fig.~\ref{fig4}), resulting in no dephasing in the limit. This discussion of the linear dispersion of $\delta_q$ is consistent with the suppression of particle-hole creation in the interaction region of the chiral channel; see Sec. \ref{Lorentzian single}.

We comment on the case of an asymmetric Mach-Zehnder interferometer, for example, where the interaction
exists only in one (saying region 3) of the two arms but the two arms have the same length of d. In the strong interaction limit of this case, the packet that propagated through either region 3 or 4 remains in the Lorentzian  form in region 5. However, the center of the packet that propagated through region 3 is located at advanced position by $d$, compared with that of the packet through region 4. The shift of the packet center by $d$ is due to the strong interaction in region 3. Hence, the visibility $\mathcal{V}$ cannot reach the maximum value of 1, and it will be suppressed. For larger $\xi/d$, the suppression is weaker, as the two packets (one moved along region 3, and the other
along 4) have more overlap in region 5. We note that in Ref. \onlinecite{Lebedev11PRL}, an asymmetric Mach-Zehnder interferometer was discussed in the context different from our study, to show that a voltage pulse is applied to undo the distortion of a single-particle wave packet due to a capacitive Coulomb interaction.


Finally, we crudely estimate the interaction parameter $g$ in experiments. One has the capacitive interaction $\sim e^2 \mathcal{N}^2 / (2 C)$, where capacitance  $C \sim \epsilon d$ and $\mathcal{N}$ is the amount of electric charges in the interferometer arm.~\cite{Roulleau08PRL} By comparing this with Eq. (\ref{interaction_Hamiltonian}), one has $g / (2 \pi) \sim e^2 / (\epsilon h v_F)$. Inserting typical experimental parameters \cite{Schneider11PRB,McClure09PRL} of $\epsilon=12.5 \epsilon_0$ and $v_F=(2 - 15) \times 10^4 \mathrm{m/s}$, one estimates $g / (2 \pi)\sim2.3 - 17.5$; $\epsilon_0=8.85 \times 10^{-12}\mathrm{C/Vm}$.
This value falls in the range where the visibility revival occurs (see Fig. \ref{fig4}). As $v_F$ or $\epsilon$ may be modulated in experiments \cite{Chung12PC} by about factor $2$, the visibility revival may be studied in experiments.



In summary, we examined the interaction effect on the coherence of a single electron wave packet of Lorentzian shape in an electronic Mach-Zehnder interferometer. In particular, we found that the visibility of the interference of the packet shows the nonmonotonic behavior as a function of interaction strength, and that in the strong interaction limit, the visibility is restored to the value of the non-interacting case. This counterintuitive result is attributed to the suppression of particle-hole excitations in the strong-interaction limit and to the fact that the packet propagates along the chiral channels. We discuss the parameter regime where one may observe the revival of the visibility in experiments.

Our study is valid and useful for the case of filling factor $\nu=1$, as it is based on the exactly solvable model for arbitrary intra-edge interaction strength and arbitrary transmission probability at the quantum point contacts. On the other hand, it does not describe the case of filling factor $\nu=2$, where inter-edge interactions play an important role. It will be interesting to investigate a combined effect of our findings and the inter-edge interactions in the filling factor $\nu=2$.

\begin{acknowledgments}
We acknowledge  Yunchul Chung, Ki-Seok Kim, Jae-Seung Jeong,  Woo-Ram Lee,  Jaeho Han, and especially D. L. Kovrizhin for useful discussions. This work was financially supported by the NRF through Grant No. 2009-0084606 (HSS) and 2011-0030784 (SYL, HWL).
\end{acknowledgments}

\appendix

\section{ $\hat{\rho}^{(1)}_l(q,t)$ and $\delta \hat{\phi}_l(x,t)$}
\label{Additional_density_and_field_operator}

In this section, we derive the expression of $\hat{\rho}^{(1)}_l(q,t)$ and {$\delta \hat{\phi}_l(x,t)$ in Eq. (\ref{delta_phiqt})}
. The commutation relation between density operators leads to the following relations
\begin{equation}
  \begin{aligned}
\,\,\, [ \hat{H}_{kin},\hat{\rho}_l(q) ] &= -q v_F \hbar \hat{\rho}_l(q),  \\
  [ \hat{H}_{int},\hat{\rho}_l(q) ] &= -\frac{g}{\pi d} v_F \hbar \sin{q d/2} \hat{\mathcal{N}}_l, \\
  [ \hat{H}_{kin},\hat{\mathcal{N}}_l ] &= i v_F \hbar \int^{d/2}_{-d/2} dx \partial_x \hat{\rho}_l(x), \\
  [\hat{H}_{kin},\int^{d/2}_{-d/2} dx
\partial^n_x \hat{\rho}_l(x)] &= iv_F \hbar \int^{d/2}_{-d/2} dx
\partial^{n+1}_x \hat{\rho}_l(x),
  \end{aligned}
  \label{[kin,rho]}
\end{equation}
where $\hat{\mathcal{N}}_l=\int^{d/2}_{-d/2} dx
\hat{\rho}_l(x)$ and $n$ is a positive integer. Using the relations, one finds
\begin{eqnarray}\label{drhoqt}\delta \hat{\rho}^{(1)}_l(q,t)&=&-\frac{g}{\pi d}v_F \hbar \sin{q d/2}
 [ \sum^{\infty}_{m=1}\sum^{\infty}_{n=m} \frac{(it/\hbar)^n}{n!} \\ &\times& (-q
v_F \hbar)^{n-m}(i v_F \hbar)^{m-1} \int^{d/2}_{-d/2} dx
\partial^{m-1} _x \hat{\rho}_l(x)]. \nonumber \end{eqnarray}
The integral in Eq.~\eqref{drhoqt} is rewritten in the terms of
$\hat{\rho}_l(q)$, the Fourier transformation of $\hat{\rho}_l(x)$, as
\be \int^{d/2} _{-d/2} dx \partial^{m-1} _x
\hat{\rho}_l(x)=\frac{d}{L} \sum_{q'} (i q')^{m-1}
\frac{\sin{q'd/2}}{q'd/2} \hat{\rho}_l(q'). \ee
To evaluate the summation in the above equation, we use
$\sum^{\infty}_{m=1}\sum^{\infty}_{n=m}=\sum^{\infty}_{m=0}\sum^{\infty}_{n=m}-\sum^{\infty}_{n=0}(m=0)=\sum^{\infty}_{n=0}\sum^{n}_{m=0}-\sum^{\infty}_{n=0}(m=0)$.
Since the total system length $L$ is much larger than $d$ and $v_F t$, a summation over $q$ is converted to an integral as $ \sum_{q\neq0} \rightarrow \int^{\infty}_{-\infty} \frac{dq}{2 \pi /L}$.
This yields $\hat{\rho}^{(0)}_l(x,t)=e^{-i q v_F t} \hat{\rho}_l (q,0)$  and
\begin{eqnarray} \label{delta_rhoqt} \delta \hat{\rho}^{(1)}_l(q,t) &=& \frac{g}{\pi d} \sin{q d/2} \frac{d}{L} [ \sum_{q'} \frac{\sin{q'd/2}}{q'd/2} 
\hat{\rho}_l(q')\nonumber\\ &\times& [\frac{1}{q-q'}(e^{-iqv_F t}-e^{-iq'v_F t})].\end{eqnarray}
By inserting Eq. (\ref{delta_rhoqt}) into Eq. (\ref{bosonicfields}), we find
\begin{eqnarray} \label{1stphi} \delta \hat{\phi}^{(1)}_l(x,t) &=& \frac{4g}{L^2}
\sum_{q'} \frac{\sin{q'd/2}}{q'd}  \hat{\rho}_l(q') \\
 & \times & \sum_{q\neq0}\frac{e^{iqx}e^{-|q|a/2}}{iq} [\frac{e^{-iqv_F t}-e^{-iq'v_F t}}{q-q'}]  \sin \frac{qd}{2}. \nonumber \end{eqnarray}

We next derive Eqs.~\eqref{equationofmotion} and \eqref{delta_phiqt}.
The equation of motion of the first order $\delta \hat{\phi}^{(1)}_l(x,t)$ is obtained from the partial derivative of the right-hand side of Eq.~\eqref{1stphi}, $[\partial_t+v_F \partial_x]\frac{1}{L} \sum_{q\neq0}\frac{1}{iq}e^{iqx}e^{-|q|a/2}\sin{q d/2}
[\frac{1}{q-q'}(e^{-iqv_F t}-e^{-iq'v_F t})]=-\frac{v_F}{4} e^{-i q'v_F t} [\textrm{sgn}(x+d/2)-\textrm{sgn}(x-d/2)] $ {when $a \rightarrow 0$}. Here, $\textrm{sgn}(x)=\pm 1$ for $x\gtrless0$ and $\textrm{sgn}(x)=0$ for $x=0$. Then, one can verify  that $ [\partial_t+v_F \partial_x] \hat{\phi}^{(1)}_l(x,t)=-v_F \frac{g}{2 \pi} [\hat{\phi}^{(0)}_l(\frac{d}{2},t)-\hat{\phi}^{(0)}_l(-\frac{d}{2},t)+\frac{d}{L}\hat{N}_l]$ for $-d/2\leq x \leq d/2$, and $[\partial_t+v_F \partial_x] \hat{\phi}^{(1)}_l(x,t)=0$ otherwise.
{Similarly,}
we obtain the recursive relation as
\be
[\partial_t+v_F \partial_x] \hat{\phi}^{(n+1)}_l(x,t)=-v_F \frac{g}{2 \pi} [\hat{\phi}^{(n)}_l(\frac{d}{2},t)-\hat{\phi}^{(n)}_l(-\frac{d}{2},t)]\ee for $n \geq 1$.
The trial solution of the $n+1$-th order for the long time limit of $t>d/v_F$ is
\begin{equation}
\label{n1stphi}
\begin{aligned}
\delta \hat{\phi}^{(n+1)}_l(x,t) &= \frac{4 g}{L^2} \sum_{q'} \frac{1}{q'd} \sin \frac{q'd}{2} (-\frac{g}{\pi}\frac{e^{iq'd/2}}{q'd} \sin \frac{q'd}{2} )^{n}  \\ &\times \hat{\rho}_l(q') \sum_{q\neq0}\frac{e^{iqx}}{iq} \sin \frac{q d}{2} [\frac{e^{-iqv_F t}-e^{-iq'v_F t}}{q-q'}].
\end{aligned}
\end{equation}
Here, we used $\frac{1}{L} \sum_{q \neq 0} \frac{1}{iq}e^{iqd/2} \sin \frac{qd}{2} \frac{1}{q-q'}(e^{-i q v_F t}-e^{-iq'v_F t}) = - \frac{1}{q'} \sin \frac{q'd}{2} e^{i q' d/2}e^{-iq'v_F t}$ {for $\hat{\phi}^{(n)}_l(d/2,t)$}, and $\hat{\phi}^{(n)}_l(-d/2,t) =0$.
By summing all the orders, we derive Eq. (\ref{delta_phiqt}).

\section{$\delta \rho_\textrm{ch}(Y,t)$ in the chiral-channel case}
\label{Additional density operator}

In this section, we derive Eq. (\ref{Additional density operator eq}).
For $Y \gg d/2$, $\delta \rho_{\textrm{ch}}(Y,t)=\langle \Psi_{\xi}(X)  | \delta\hat{\rho}(Y,t) |\Psi_{\xi}(X)\rangle$ is written as
\bea \delta\rho_{\textrm{ch}}&=&\frac{2 \xi}{L}\sum_{k,k'>0} -i(k'-k) K(k'-k) \nonumber\\ & &\times e^{i (k'-k)v_F(X+Y-v_F t)} e^{-(k+k')\xi}.\eea
By putting $q=k'-k$ and $Q=k'+k$,
we obtain
\be \delta\rho_{\textrm{ch}}=
\frac{2 \xi}{L} \sum_{q} -iq K(q) e^{i q(X+Y-v_F t)}\sum_{Q>|q|} e^{-Q\xi}. \ee
Notice that the discrete unit of $Q$ is $2 \Delta k= 4 \pi / L$.
This relation is reduced to Eq. (\ref{Additional density operator eq}),
after the summation over $Q$ is performed in the limit of  $L\rightarrow\infty$.

\section{$\rho_o (Y, t)$ in the interferometer case}
\label{bosoncorrelator}

In this section, we derive $\rho_o (Y, t)$  in Eq. (\ref{densityprofileo}).
$\chi_l(Y,t)$ in Eq.(\ref{chi}) is reexpressed in terms of bosonic field operators in Eq.(\ref{fermion-boson-relation}) as
\begin{equation}
\label{bosonizedchi} \chi_l(Y,t)= \int dx'\frac{1}{2 \pi a} \langle F| e^{i \hat{\phi}_{l}(Y,t)} e^{-i \hat{\phi}_{l,0}(x',0)}|F \rangle f_{\xi}(x';X).
\end{equation}
$\hat{\phi}_{l,0}(x',0)$ stands for the bosonic field of the non-interacting case, satisfying $\hat{\phi}_{l,0}(x',0)=\hat{\phi}_{l,0}(x')=\frac{2 \pi}{L}\sum_{q\neq 0} \frac{1}{iq} e^{iqx'} e^{-|q|a/2} \hat{\rho}_l(q)$. $\hat{\phi}_{l}(Y,t)$ is decomposed to $\hat{\phi}_{l}(z)=\hat{\phi}_{l,0}(z)+\delta \hat{\phi}_{l}(z)$ with $z=Y-v_Ft$ where $\delta \hat{\phi}_{l}(z)=-\sum_{q\neq 0} K(q)e^{iqz} \hat{\rho}_l(q)$ for $Y \gg d/2$ (regions 5, 6) and $t>\frac{x+d/2}{v_F}$ (propagation time from $x=-d/2$ to $x$). $\hat{\phi}_{l,0}(x')$ can be divided into an annihilation operator and a creation operator of the bosonic field, $\hat{\varphi}_{l,0}(x')\equiv\frac{2 \pi}{L}\sum_{q>0} \frac{1}{iq} e^{iq(x'+ia/2)} \hat{\rho}_l(q)$ and $\hat{\varphi}^{\dagger}_{l,0}(x')\equiv \frac{2 \pi}{L} \sum_{q<0} \frac{1}{iq} e^{iq(x'-ia/2)} \hat{\rho}_l(q)$. Similarly, $\delta\hat{\phi}_l(z)$ is also divided into $\delta\hat{\varphi}_l(z)\equiv-\sum_{q>0} K(q) e^{iqz} \hat{\rho}_l(q)$ and $\delta\hat{\varphi}^{\dagger}_l(z)\equiv-\sum_{q<0} K(q) e^{iqz} \hat{\rho}_l(q)$.  Since $\hat{\varphi}_{l,0}(x')|F \rangle=\hat{\varphi}_l(z)|F \rangle=\langle F  |\hat{\varphi}^{\dagger}_{l,0}(x')=\langle F  |\hat{\varphi}^{\dagger}_{l}(z)=0$, one can move annihilation (creation) operators to the right (left) side in Eq.~\eqref{bosonizedchi}.  Using $e^{(A+B)}=e^A e^B e^{-\frac{1}{2} [A,B]}$ and $e^A e^B= e^B e^A e^{[A,B]}$, we find that the integrand of Eq.~\eqref{bosonizedchi} has the form of
\begin{eqnarray} \frac{1}{2 \pi a}e^{\frac{1}{2}[\hat{\varphi}^{\dagger}_{l}(z), \hat{\varphi}_{l}(z)]}e^{\frac{1}{2}[\hat{\varphi}^{\dagger}_{l,0}(x'), \hat{\varphi}_{l,0}(x')]} e^{[\hat{\varphi}_{l}(z), \hat{\varphi}^{\dagger}_{l,0}(x')]}.
\nonumber
\end{eqnarray}
We compute $ [\hat{\varphi}^{\dagger}_{l,0}(z), \hat{\varphi}_{l,0}(z)]=[\hat{\varphi}^{\dagger}_{l}(x'), \hat{\varphi}_{l}(x')]=\log \frac{2 \pi a}{L}$, by using $\sum_{n>0}\frac{\exp({-2 n\pi a/L})}{n}= - \log(1-e^{-\frac{2 \pi a}{L}})$.
Similarly, $ [\hat{\varphi}_{l}(z), \hat{\varphi}^{\dagger}_{l,0}(x')]=[\hat{\varphi}_{l,0}(z), \hat{\varphi}^{\dagger}_{l,0}(x')]+[\delta\hat{\varphi}_{l}(z), \hat{\varphi}^{\dagger}_{l,0}(x')]$,
$ \exp ([\hat{\varphi}_{l,0}(z), \hat{\varphi}^{\dagger}_{l,0}(x')]) \simeq i L / [ 2 \pi (z-x'+ia)]$ in the limit of $L \to \infty$,
and $[\delta\hat{\varphi}_{l}(z), \hat{\varphi}^{\dagger}_{l,0}(x')]=-i \sum_{q>0} K(q) e^{i q (z-x'+i a/2)}$. The last relation captures interaction effects.

Then, we compute the integral of Eq.~\eqref{bosonizedchi}, by using the contour integration of complex variable $x' \to z'$,
\bea \chi_l(z) =  \sqrt{\frac{\xi}{\pi}}\oint dz' \frac{i e^{-i \sum_{q>0} K(q) e^{i q (z-z'+ia/2)}} }{2 \pi (z'-z-ia)(z'-X+i \xi)}.
\nonumber \eea
One pole exists at $z'=z+ia$ 
in the upper plane. 
And another at $z'=X-i\xi$ in the lower plane.
By choosing the lower-plane contour including the pole at $z'=X-i\xi$, we obtain
 \be \chi_l(z)= \mathrm{phase} \times \sqrt{\frac{\xi}{\pi}}\frac{1}{z-X+i \xi} e^{\mathrm{Im}[\sum_{q>0}  K(q) e^{iq(z-X)}e^{-q \xi}]}, \ee where $a\rightarrow 0$. Since $\chi_u(z)=\chi_d(z)=\chi(z)$ in the symmetric case, $\rho_o(Y,t)=|\chi(z)|^2$. After evaluating $|\chi(z)|^2$, we derive Eq. (\ref{densityprofileo}).

\bibliographystyle{apsrev}


\end{document}